\documentclass[twocolumn,showpacs,showkeys]{revtex4}
\usepackage{amsmath}
\usepackage{amsfonts}
\usepackage{amssymb}
\usepackage{graphicx}

\newcommand{\wb}{\omega_{\mathrm{b}}}
\newcommand{\wo}{\omega_{\mathrm{o}}}
\newcommand{\wn}{\omega_{\mathrm{n}}}
\newcommand{\wL}{\omega_{\mathrm{L}}}

\newcommand{\singlefig}{.375\textwidth}

\begin{document}

\title{Moving discrete breathers in a Klein--Gordon chain with an impurity}
\author{J Cuevas}
\affiliation{%
 ETS Ingenier\'{\i}a Inform\'atica. Universidad de Sevilla. Avda Reina Mercedes s/n, 41012-Sevilla,
 Spain}
\author{F Palmero}
\affiliation{%
 ETS Ingenier\'{\i}a Inform\'atica. Universidad de Sevilla. Avda Reina Mercedes s/n, 41012-Sevilla,
 Spain}
\author{JFR Archilla}
\affiliation{%
 ETS Ingenier\'{\i}a Inform\'atica. Universidad de Sevilla. Avda Reina Mercedes s/n, 41012-Sevilla,
 Spain}
\author{FR Romero}
\affiliation{%
 Facultad de F\'{\i}sica. Universidad de Sevilla. Avda Reina Mercedes s/n, 41012-Sevilla,
 Spain}

\begin{abstract}
We analyze the influence of an impurity in the movement of
discrete breathers in Klein--Gordon chains. We observe that the
moving breather can cross the impurity, can be reflected by it, or
can be trapped originating a quasi-periodic breather. We find that
resonance with a nonlinear localised mode centred in the impurity
is a necessary condition in order to observe the trapping
phenomenon, as a difference with the resonance condition with a
linear localised mode when this problem is studied within the
Nonlinear Schr\"{o}dinger Equation approximation.

\end{abstract}



\maketitle

\section{Introduction}

A nonlinear phenomena which has been paid a lot of attention
during recent years concerns to the localisation of oscillations
in discrete nonlinear Klein--Gordon lattices. Since the discovery
of intrinsic localised modes or \emph{discrete breathers} by
Sievers and Takeno \cite{ST88}, they have been found in a great
number of systems. MacKay and Aubry \cite{MA94} proved the
existence of discrete breathers in Klein--Gordon lattices and
established a method for calculating them as exact solutions of
the dynamical equations (for a review, see \cite{FW98,A97}). In
addition, these localized oscillations, under certain conditions,
can move and transport energy  and they are usually called
\emph{moving breathers} \cite{CAT96,AC98,CAGR02,IST02,CPAR02}.

A method used for calculating discrete breathers in Klein--Gordon
lattices consists in approximating the dynamical equations by a
Nonlinear Schr\"{o}dinger Equation (NLS) and the moving breather
by an envelope soliton. One problem that had been studied using
the NLS limit concerned to the interaction of moving breathers
with an impurity  supposing that the breather has a small
amplitude \cite{FPM94}. However, moving breathers involve
oscillations of large amplitude, which implies that the NLS
approximation may not be as accurate as expected. The aim of this
paper is to study the influence of an impurity in the movement of
discrete breathers in a Klein--Gordon chain, i.e. using the
dynamical equations without any approximation, a problem which has
not been yet undertaken.

In Klein--Gordon lattices, the local modes due to an impurity and
the localized modes due to the nonlinearity of the lattice can be
considered equivalent entities \cite{AMM99,CAPR01}. This fact
suggests that a moving breather can exhibit an interesting
behaviour when interacts with an impurity. As a matter of fact, we
have found that the moving breather can cross the impurity, can be
reflected by it (being able to let the impurity excited), or can
be trapped, generating a local depository of energy. The results
obtained in our study show that it is necessary the resonance of
the moving breather with a nonlinear local mode (i.e. a discrete
static breather centred in the impurity site) in order to generate
a trapped breather.

This behaviour is qualitatively similar to the observed in the NLS
approximation \cite{FPM94}. However, the ranges of the values for
which the different phenomena occur are rather different.
Furthermore, the most remarkable difference between their
approximation and our results is that in they obtained that the
trapping is due to a resonance with a linear local mode, whereas
in our case, the resonance must involve a nonlinear local mode.

In addition, we have observed the non existence of trapping
phenomena even though resonance with the nonlinear local mode
occur. We have also observed that in this case the tails of the
nonlinear local mode and the lineal local mode have different
vibration patterns. As a consequence, we conjecture that both
tails must have the same vibration pattern in order that the
trapping occur.

\section{Model and solutions generation} \label{model}

\subsection{Formulation of the model}

In order to study the effects of impurities on breather mobility,
we consider a simple model where moving breathers can be
generated, that is, a Klein--Gordon chain with nearest neighbours
attractive interaction \cite{CAT96,AC98}, whose Hamiltonian is
given by:

\begin{equation}\label{ham}
    H=\sum_n\left(\frac{1}{2}\dot
    u_n^2+V_n(u_n)+\frac{1}{2}C(u_n-u_{n-1})^2\right),
\end{equation}
where $u_n$ represents the displacement of the n-th particle with
respect to its equilibrium position, $C$ is a coupling constant
and $V_n(u_n)$ is the on--site potential at the n-th site. We
choose $V$ as the Morse potential, i.e., $V_n(u)=D_n(e^{-u}-1)^2$,
which proves to be a very suitable one to obtain moving breathers
\cite{AC98,CAGR02,CPAR02}. $D_n$ represents the well depth in the
n-th site. An impurity will be introduced in the chain by means of
an inhomogeneity in the 0-th well, i.e.,
$D_n=D_o(1+\alpha\delta_{n,0})$, with $\alpha$ being a parameter
which gives account of the magnitude of the impurity. It takes its
values in the interval $[-1,\infty)$. Hereafter, we will consider
$D_o=1/2$. Notice that the impurity can be also introduced in the
coupling. We have checked that the results are qualitatively the
same and will be published elsewhere.

This kind of chain has been used as a simple model of DNA
dynamics; in that context it is usually referred to as
Peyrard-Bishop model \cite{PB89}. In the framework of this model,
the variables $u_n$ represents the transverse stretching of the
hydrogen bonds connecting the two bases, $D$ is the dissociation
energy of a base pair, and $C$ is the stacking coupling constant.

This Hamiltonian leads to the dynamical equations
\begin{equation}\label{dyn}
    F(\{u_n\})\equiv\ddot{u}_n+V'_n(u_n)+C(2u_n-u_{n+1}-u_{n-1})=0.
\end{equation}

These equations have two kinds of solutions, linear ones, which
correspond to oscillations of small amplitude, and nonlinear ones,
which correspond to intrinsic localised modes (discrete
breathers).

\subsection{Linear modes}

The dynamical equations can be linearised if the amplitude of the
oscillations is small. Thus, the equations (\ref{dyn}) can be
written as:
\begin{equation}\label{linear}
    \ddot{u}_n+\wn^2u_n+C(2u_n-u_{n+1}-u_{n-1})=0,
\end{equation}
where $\wn$ is the natural frequency of the n-th oscillator in the
harmonic limit. It is related to the well depth of the Morse
potential in the form $\wn^2=2D_n$, which implies that
$\wn^2=\wo^2(1+\alpha\delta_{n,0})$, with $\wo=1$ as $D_o$ has
been chosen to be $1/2$.

Thus, this equation has $N-1$ solutions (being $N$ the number of
particles) corresponding to linear extended modes (LEMs) and one
localised solution, which corresponds to a linear local mode (LLM)
whose origin lies in the introduction of the impurity.

The frequencies of the LEMs can be calculated supposing that they
are plane waves ($u_n(t)=e^{i\omega(q)t-nq}$) and that the LLM
decays in the space following a dependence of the form
$u_n(t)=e^{i\wL t}r^n$ \cite{FPM94}. Thus, the frequencies of the
LEMs are given by:
\begin{equation}\label{LEMs}
    \omega(q)=\sqrt{\omega_o^2+4C\sin^2\frac{q}{2}},
\end{equation}
where $q$ is the LEM wave vector and takes its value in the
interval $(0,\pi]$ if $\alpha<0$ and in $[0,\pi)$ if $\alpha>0$.

The frequency of the LLM is given by the relation \cite{FPM94}:
\begin{equation}\label{LLM}
    \omega^2_{L}=\wo^2+2C+sgn(\alpha)\sqrt{\alpha^2+4C^2}.
\end{equation}

\begin{figure}
\begin{center}
\includegraphics[width=\singlefig]{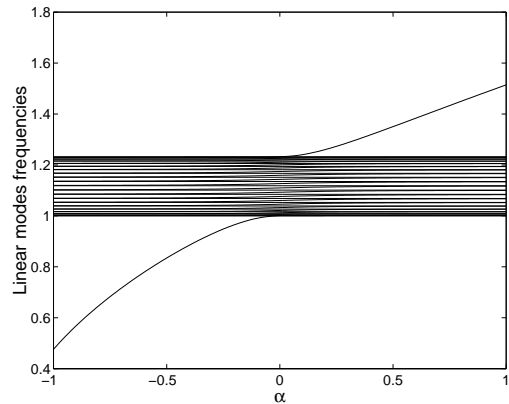}
\caption{Frequencies of the nonlinear modes in function of
parameter $\alpha$. They correspond to $C=0.13$ although the
dependence is qualitatively similar for every value of $C$.
}\label{linearmodes}
\end{center}
\end{figure}

It is worth remarking that LLMs with $\alpha<0$ have wave vector
$q=\pi$ (the particles vibrate in zigzag) while the solution for
$\alpha>0$ corresponds to $q=0$ (the particles vibrate in phase).
Figure \ref{linearmodes} shows the dependence of the frequencies
of the linear modes with $\alpha$, where the isolated frequencies
corresponds to the LLMs.

The knowledge of the frequency of the LLM is very important in
order to explain the properties of moving breathers, because
discrete breathers and local modes can be viewed as equivalent
entities when the interplay between nonlinearity and inhomogeneity
is considered \cite{AMM99,CAPR01}.

\subsection{Static and moving breathers}

A static breather can be obtained solving the full dynamical
equations. It can be done using common methods based on the
anticontinuous limit \cite{MA96}. The implementation of these
methods basically consists in calculating the orbit of an isolated
oscillator at a fixed frequency $\wb$, and using this solution as
a seed for solving the complete dynamical equations by means of a
Newton--Raphson continuation method.

If the oscillator initially chosen is the corresponding to the
impurity, a static breather centred in the impurity is obtained.
It will be called \emph{nonlinear local mode} (NLLM), in analogy
to the linear local modes (LLMs) that appear in the spectrum of
linear modes (equation \ref{LLM}).

Once a static breather is calculated, it can be moved under
certain conditions. There exists a systematic method to calculate
moving solutions \cite{CAT96,AC98} which consists in adding to
the velocities of the static breather a perturbation of magnitude
$\lambda$ colinear to the direction of the pinning mode, and
letting the system evolve in time. This perturbation breaks the
shift translational symmetry of the system. In addition, the
coupling of the static breather must be strong enough
\cite{CAGR02}.

In this paper, we have studied breathers with a frequency
$\wb=0.8$ and a coupling $C=0.13$ although this value has been
increased in order to make some comparisons. These values of the
parameters provide moving breathers with low phonon radiation for
values of the perturbation $\lambda\lesssim0.2$. We have
considered different impurities given by different values of
parameter $\alpha$ in the range $-1 \le \alpha \le 1$.

\section{Interaction of moving breathers with impurities} \label{moving}

\subsection{Numerical observations}

We have studied the behaviour of moving breathers when they
interact with an impurity. This study is performed varying the
value of the magnitude of the impurity $\alpha$, and shows an
interesting behaviour. We have found some critical values of
parameter $\alpha$ ($\alpha_1<\alpha_2<\alpha_3<0$) which separate
the different regimes (see figure \ref{range}):

\begin{figure}
\begin{center}
\includegraphics[width=\singlefig]{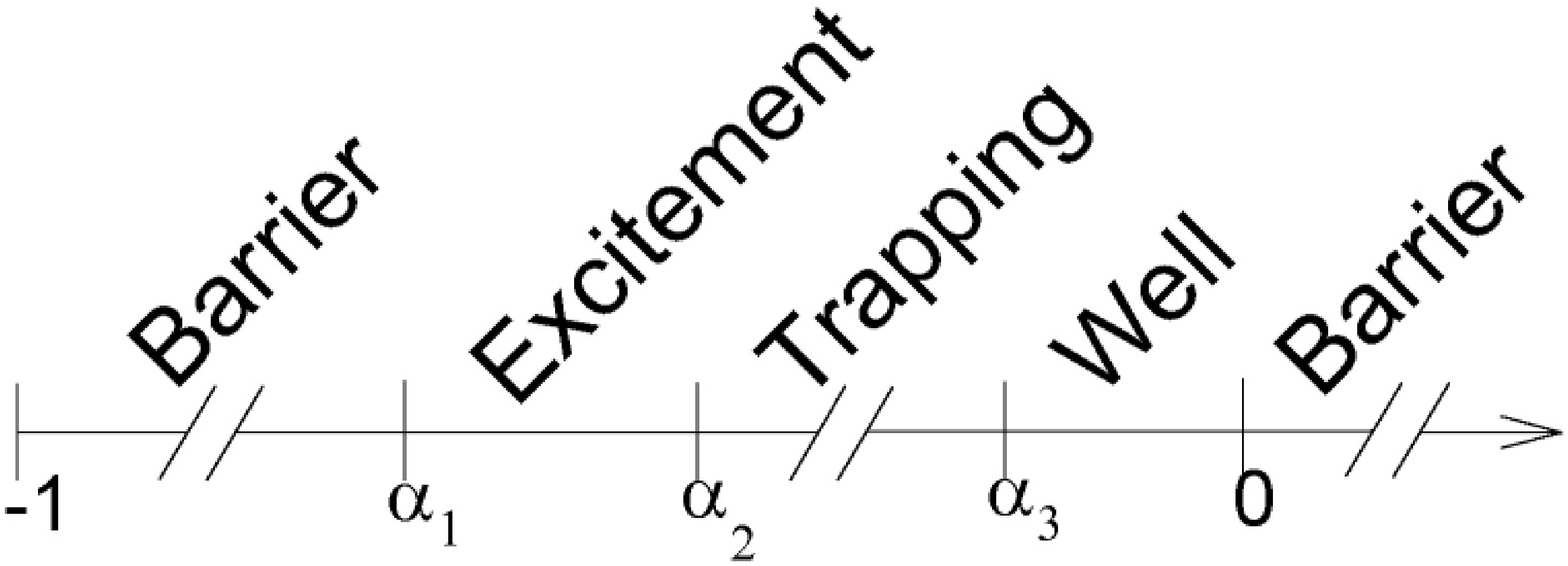}
\caption{Different regimes in the interaction of the moving
breather and the impurity (see text).}\label{range}
\end{center}
\end{figure}
\begin{enumerate}

\item
\emph{The impurity acts as a potential barrier}. It occurs
whenever $\alpha>0$ and $\alpha\in(-1,\alpha_1)$ with
$\alpha_1<0$. The breather rebounds when reaches the impurity and
let it excited during a brief time lapse. The amplitude of this
excitement decreases when the impurity is higher, i.e., when
$|\alpha|$ increases.  Note that if $\alpha\approx0$, the
potential barrier can be observed, i.e., the breather can cross
the impurity provided the translational velocity is high enough,
as it occurred in \cite{CPAR02}.

\item
\emph{The impurity is excited and the breather rebounds}. It
occurs for $\alpha\in(\alpha_1,\alpha_2)$. The energy of the
excited impurity is higher than the energy of the NLLM for the
considered value of $\alpha$ and whose frequency is the same as
the incident moving breather, $\wb$. It implies that the excited
impurity will vibrate with a frequency smaller than $\wb$ because
the on--site potential is soft. This behavior is shown in figure
\ref{excitation}.

\begin{figure}
    \includegraphics[width=\singlefig]{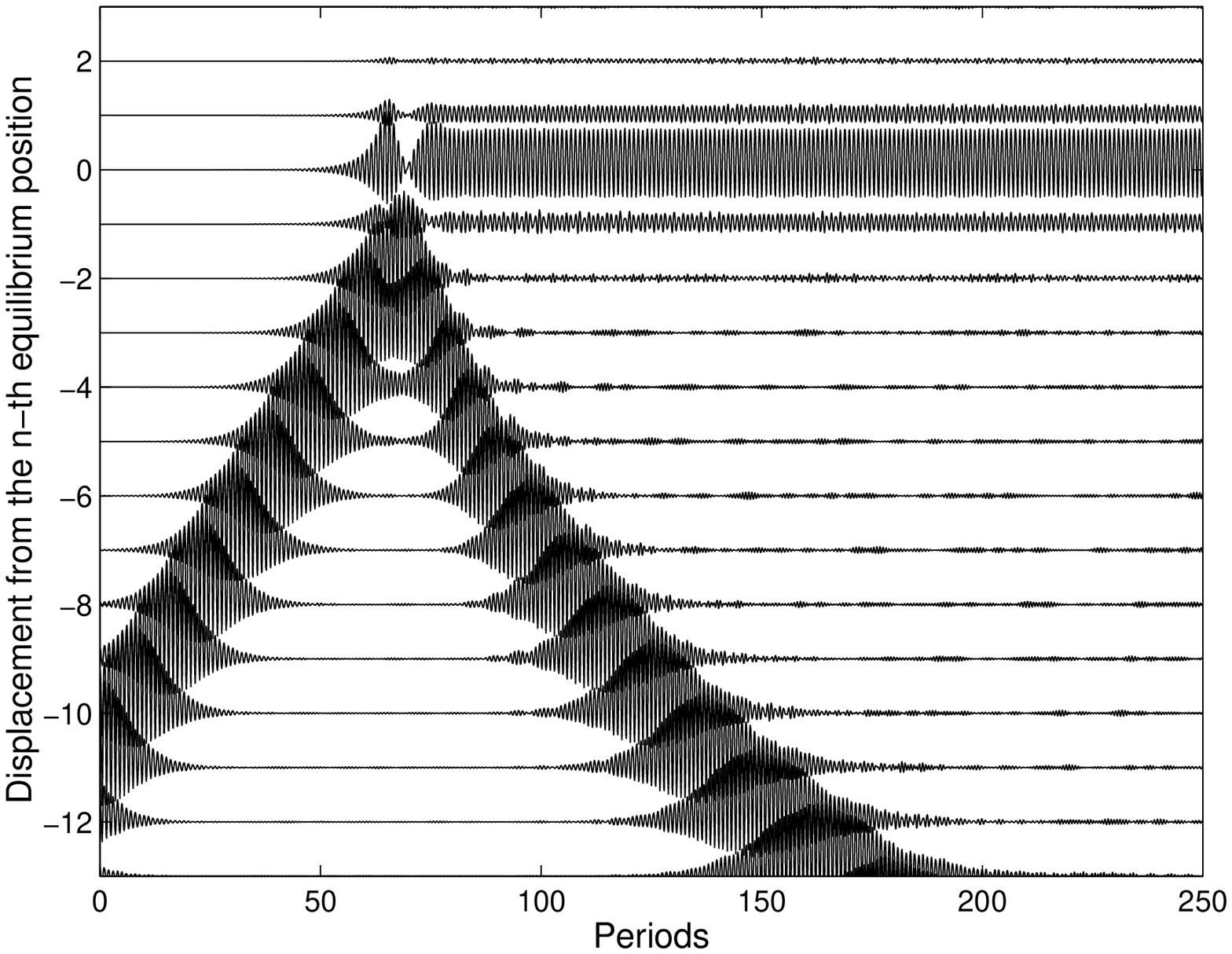} \\
    \includegraphics[width=\singlefig]{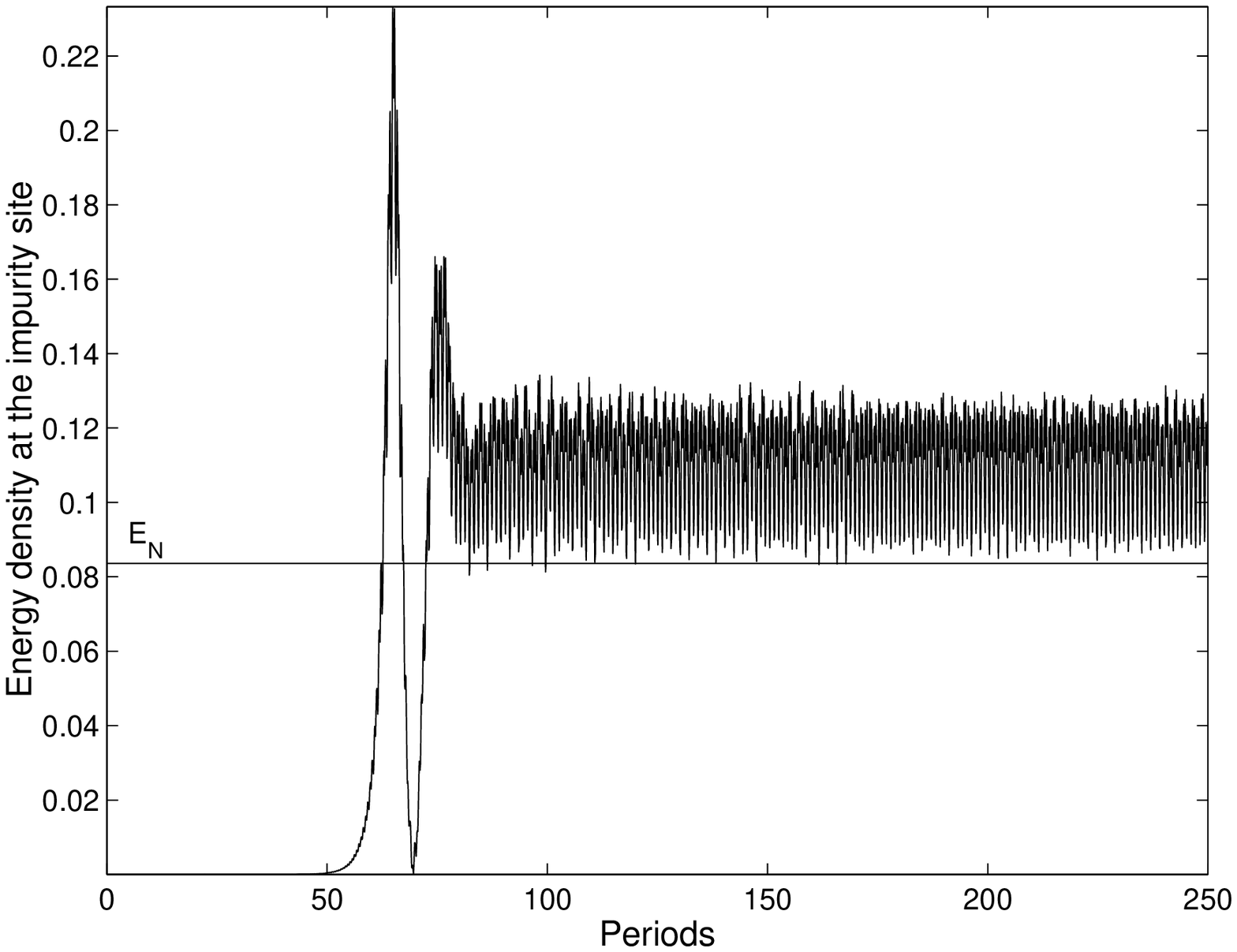}
\caption{Interaction of the breather with the impurity for
$\alpha=-0.52$ and $\lambda=0.1$. Top: Evolution of the moving
breather. Bottom: Evolution of the averaged energy density of the
impurity. $E_N$ is the energy of the NLLM.}\label{excitation}
\end{figure}

\item
\emph{The breather is trapped by the impurity}. It occurs in the
interval $\alpha\in(\alpha_2,\alpha_3)$. When the breather is near
the impurity, it is absorbed and remains trapped while its center
oscillates between the neighbouring sites. The trapping of the
breather is shown in figure \ref{trapping}. Furthermore, the
trapped breather is quasi-periodic, as can be deduced from its
Fourier spectrum (Figure \ref{fourtrapping}), and, as a
consequence, it emits a great amount of phonon radiation.

\begin{figure}
\begin{center}
\includegraphics[width=\singlefig]{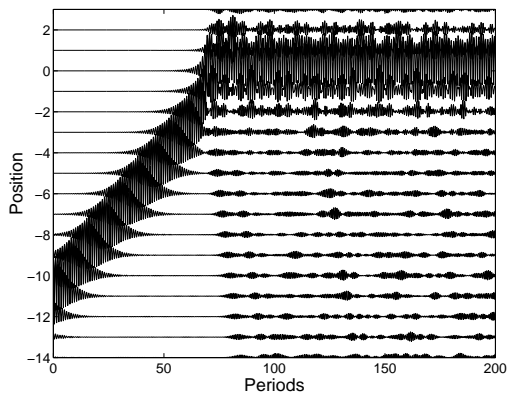}
\caption{Evolution of the moving breather for $\alpha=-0.3$ and
$\lambda=0.1$. It can be observed that the breather is absorbed by
the impurity and gets trapped; afterwards, the breather emits
phonon radiation and its energy centre oscillates between the
sites adjacent to the impurity, }\label{trapping}
\end{center}
\end{figure}

\begin{figure}
\begin{center}
\includegraphics[width=\singlefig]{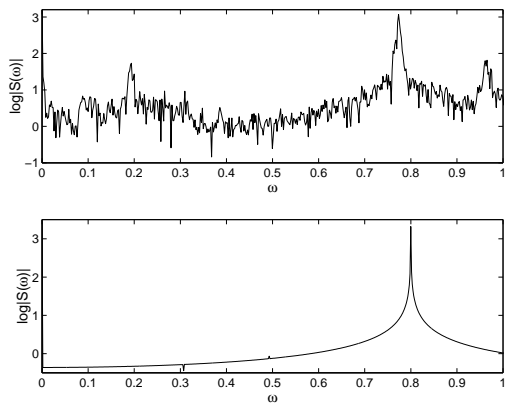}
\caption{Fourier spectra for the trapped breather of figure
\ref{trapping} (top) and a static breather with the same values of
$C$ and $\wb$. It can be observed the existence of only one peak
in the last case where there exist a great number of peaks in the
first case, which is an evidence of quasi-periodicity.}
\label{fourtrapping}
\end{center}
\end{figure}

\item
\emph{The impurity acts as a potential well}. It occurs when
$\alpha_3<\alpha<0$ and manifests as an acceleration of the
breather when reaches a site near the impurity, and a deceleration
when the impurity is crossed.

\end{enumerate}

All of these cases have been studied using a breather with
$\wb=0.8$ and $C=0.13$. The translational velocity of the moving
breathers has been chosen small so as not to have high phonon
radiation. The critical values of $\alpha$ cannot be determined
because the transition between the different behaviours are
diffuse. An estimation of the critical values for $\wb=0.8$ and
$C=0.13$ are: $\alpha_1\approx-0.54$, $\alpha_2\approx-0.49$ and
$\alpha_3\approx-0.02$. These regimes have also been found for
different values of the frequency, although the value of
$\alpha_2$ decreases with $\wb$.

If the coupling is higher, the phonon radiation is not so low and
can mask some of the effects.

\subsection{Justification of some results}

\begin{figure}
\begin{center}
\includegraphics[width=\singlefig]{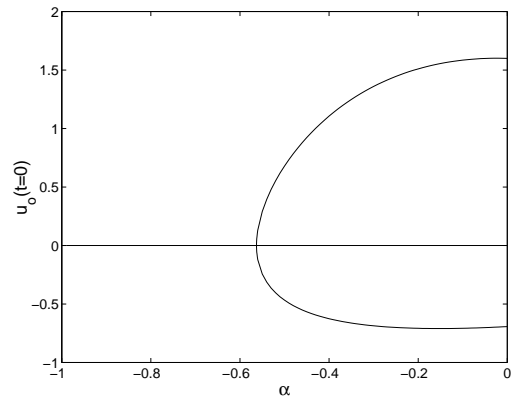}
\caption{Pitchfork bifurcation for $\wb=0.8$ and $C=0.13$.
$\alpha_{res}=-0.5328$.}\label{pitchfork}
\end{center}
\end{figure}

Some of the results exposed in the last subsection can be
explained from the properties of the NLLMs. Concretely, if a
continuation of the static breathers is performed varying the
parameter $\alpha$, a bifurcation appears for
$\alpha\equiv\alpha_c>0$ and another one for
$\alpha\equiv\alpha_{res}<0$. The first one is originated by a
localized Floquet eigenmode which abandons the unit circle. In the
second case, the breather bifurcates with the stationary solution
through a pitchfork (figure \ref{pitchfork}). In this bifurcation,
the outer branches correspond to NLLMs either with $u_n>0$ for
$t=0$ and with $u_n<0$ for $t=0$, while the central branch
corresponds to the stationary solution, i.e. all the oscillators
at rest. This kind of bifurcation is different to the broken
pitchforks obtained in disordered systems
\cite{AMM99,CAPR01,KA99}. It is due to the fact that, even thought
an isolated impurity breaks the shift translational symmetry of
the system, it does not break the mirror symmetry. Consequently,
the pitchforks are not broken.
The value $\alpha_{res}$ coincides with the point at which the
frequency of a LLM is the same as the frequency of the NLLM, i.e.,
$\wL=\wb$. This value can be calculated from the equation
\ref{LLM} as a function of $\wb$ and $C$:

\begin{equation}\label{resonance}
    \alpha_{res}=-\sqrt{(\wb^2-\wo^2)(\wb^2-\wo^2-4C)}
\end{equation}

Thus, for $C=0.13$ and $\wb=0.8$, $\alpha_{res}=-0.5628$ which is
lower than $\alpha_1$. We have also checked for different values
of the frequency and coupling and we have found that there are
situations where $\alpha_1$ is slightly smaller that
$\alpha_{res}$ but always $\alpha_{res}<\alpha_2$. However, the
trapped breather does not exist for $\alpha>0$. It indicates that
$\alpha\in(\alpha_{res},0)$ is a necessary condition for the
existence of the trapped breather.

A conjecture for the existence of trapped breathers when
$\alpha<0$ is the following: in this case, the LLM has $q=0$, and
also all the particles of the NLLM vibrates in phase; this
vibration pattern indicates that the NLLM bifurcates from plane
waves with $q=0$ \cite{F96}, i.e., the NLLM bifurcates from the
LLM and it will be the only localized mode that exists when the
impurity is excited for $\alpha>\alpha_{res}$. In fact, we have
performed successful a continuation from the NLLM to the LLM at
constant action and disorder \cite{AMM99,CAPR01}. Thus, when the
moving breather reaches the impurity, it can excite the NLLM.
However, Doppler effect makes the NLLM `see' the moving breather
with a frequency different from $\wb$ so that when they interact,
they merge into an entity (the trapped breather) with two
frequencies.

When $\alpha_{res}<\alpha<\alpha_2$, the NLLM is unable to create
a trapped entity. It is observed that the energy of the NLLM
decreases with $|\alpha|$, so, there must be threshold for the
energy of the NLLM in order that the trapped breather can exist.
The narrow window of impurity excitements observed in the interval
$(\alpha_1,\alpha_2)$ can be due to a resonance of the moving
breather with the NLLM of frequency slightly smaller that $\wb$
and, therefore, the excited impurity has a higher energy than the
NLLM of frequency $\wb$, as we are considering a soft potential.

For $\alpha<\alpha_{res}$, the trapped breather cannot be
generated, and the moving breather always rebounds. In addition,
the NLLM does not exist. Therefore, there must be connection
between both facts, i.e., the existence of the NLLM will be a
necessary condition in order to obtain a trapped breather.

If $\alpha>0$, the scenario is different. In this case, the LLM
has $q=\pi$ but the NLLM's particle vibrate again in phase, that
is, the NLLM will not bifurcate from the LLM. Thus, there are two
different localized modes for $\alpha>0$: the LLM and the NLLM.
But, actually, the equations that govern the system are nonlinear,
so the linear modes can only correspond to low-amplitude
oscillations. In the case of the NLLM, the linear regime
corresponds to the tails. Thus, if the moving breather reaches the
impurity site, it will excite the NLLM, its tails and also the
tails of the LLM. But the latter vibrate in zigzag. As a
consequence, there will be two different linear localized
entities: the tails of the LLM (vibrating in zigzag) and the tails
of the NLLM (vibrating in phase). Therefore, we conjecture that
the existence of both linear localized entities at the same time
may be the reason why the impurity is unable to trap the breather
when $\alpha>0$.

To summarize, the existence of a NLLM for a certain value of
$\alpha$ is necessary for the existence of trapped breathers.
However, if there exists a LLM with a different vibration pattern
to the NLLM, the trapped breather does not exist.

\section{Conclusions}

We have studied the interaction of a moving discrete breather in a
Klein--Gordon chain with an impurity at rest. A rich behaviour is
observed when the breather reaches the impurity. It can be
summarized as follows: a) the impurity can act as a potential
barrier or a potential well; b) the breather can be reflected by
the impurity which is let excited; c) the impurity can trap the
breather, making it quasi-periodic.

A previous work made using the NLS approximation \cite{FPM94}
showed that moving breathers interacting with an impurity have
qualitatively the same behaviour as the observed by us. There is,
however, a very important difference between this work and ours.
It relies in the fact that in the NLS approximation, a resonance
between the frequencies of the moving breather and the linear
local mode created by the impurity is needed whereas in our case,
this resonance prevents the breather from being trapped. In our
case, the trapping phenomena is owed to the resonance between the
moving breather and a nonlinear localised mode.

The model used in our study is the same as the studied by Peyrard
and Bishop to explain DNA denaturation \cite{PB89}. It indicates
that our results can be applied to study some properties of DNA
chains. For instance, if a moving breather is generated in a DNA
chain, it can be trapped by an impurity, and act as a precursor of
the transcription bubble \cite{TP96}. This trapped breathers can
interact with another breathers and collect energy \cite{BP96}.

The impurity in DNA can have two different origins. The first one
consists in a modification of the well depth due to the action of
a transcription enzyme through a chemical effect, as explained in
\cite{TP96}. The second one relies in the fact that the A-T base
pairs have 2 hydrogen bonds whereas the C-G base pairs have 3
hydrogen bonds. So, it can be supposed that in the first case, the
well depth of the Morse potential is 2/3 of the second case. From
the results of this paper, trapping does not occur when the well
depth in the impurity is higher than in the homogeneous area. It
implies that trapping can occur in a chain of C-G pairs with an
impurity of A-T. Nevertheless, the main fault of these mechanisms
is the assumption of homogeneity in DNA, where the role of the
inhomogeneity due to the genetic code is known to be crucial in
the dynamics \cite{SK94}. However, the study of the effects of
impurities is the first step to understand the dynamics of moving
breathers in DNA sequences.

\section*{Acknowledgments}
  This work has been supported by the European Commission under
  the RTN project LOCNET, HPRN-CT-1999-00163. We acknowledge
  PG Kevrekidis for his useful comments. JC
  acknowledges an FPDI grant from `La Junta de Andaluc\'{\i}a'.

\newcommand{\noopsort}[1]{} \newcommand{\printfirst}[2]{#1}
  \newcommand{\singleletter}[1]{#1} \newcommand{\switchargs}[2]{#2#1}


\begin{thebibliography}{19}
\expandafter\ifx\csname
natexlab\endcsname\relax\def\natexlab#1{#1}\fi
\expandafter\ifx\csname bibnamefont\endcsname\relax
  \def\bibnamefont#1{#1}\fi
\expandafter\ifx\csname bibfnamefont\endcsname\relax
  \def\bibfnamefont#1{#1}\fi
\expandafter\ifx\csname citenamefont\endcsname\relax
  \def\citenamefont#1{#1}\fi
\expandafter\ifx\csname url\endcsname\relax
  \def\url#1{\texttt{#1}}\fi
\expandafter\ifx\csname
urlprefix\endcsname\relax\def\urlprefix{URL }\fi
\providecommand{\bibinfo}[2]{#2}
\providecommand{\eprint}[2][]{\url{#2}}

\bibitem[{\citenamefont{Sievers and Takeno}(1988)}]{ST88}
\bibinfo{author}{\bibfnamefont{A.}~\bibnamefont{Sievers}} \bibnamefont{and}
  \bibinfo{author}{\bibfnamefont{S.}~\bibnamefont{Takeno}},
  \bibinfo{journal}{Phys. Rev. Lett.} \textbf{\bibinfo{volume}{61}},
  \bibinfo{pages}{970} (\bibinfo{year}{1988}).

\bibitem[{\citenamefont{MacKay and Aubry}(1994)}]{MA94}
\bibinfo{author}{\bibfnamefont{R.}~\bibnamefont{MacKay}} \bibnamefont{and}
  \bibinfo{author}{\bibfnamefont{S.}~\bibnamefont{Aubry}},
  \bibinfo{journal}{\mbox{Nonlinearity}} \textbf{\bibinfo{volume}{7}},
  \bibinfo{pages}{1623} (\bibinfo{year}{1994}).

\bibitem[{\citenamefont{Flach and Willis}(1998)}]{FW98}
\bibinfo{author}{\bibfnamefont{S.}~\bibnamefont{Flach}} \bibnamefont{and}
  \bibinfo{author}{\bibfnamefont{C.}~\bibnamefont{Willis}},
  \bibinfo{journal}{Physics Reports} \textbf{\bibinfo{volume}{295}},
  \bibinfo{pages}{181} (\bibinfo{year}{1998}).

\bibitem[{\citenamefont{Aubry}(1997)}]{A97}
\bibinfo{author}{\bibfnamefont{S.}~\bibnamefont{Aubry}},
  \bibinfo{journal}{\mbox{Physica D}} \textbf{\bibinfo{volume}{103}},
  \bibinfo{pages}{201} (\bibinfo{year}{1997}).

\bibitem[{\citenamefont{Chen et~al.}(1996)\citenamefont{Chen, Aubry, and
  Tsironis}}]{CAT96}
\bibinfo{author}{\bibfnamefont{D.}~\bibnamefont{Chen}},
  \bibinfo{author}{\bibfnamefont{S.}~\bibnamefont{Aubry}}, \bibnamefont{and}
  \bibinfo{author}{\bibfnamefont{G.}~\bibnamefont{Tsironis}},
  \bibinfo{journal}{Phys. Rev. Lett.} \textbf{\bibinfo{volume}{77}},
  \bibinfo{pages}{4776} (\bibinfo{year}{1996}).

\bibitem[{\citenamefont{Aubry and Cretegny}(1998)}]{AC98}
\bibinfo{author}{\bibfnamefont{S.}~\bibnamefont{Aubry}} \bibnamefont{and}
  \bibinfo{author}{\bibfnamefont{T.}~\bibnamefont{Cretegny}},
  \bibinfo{journal}{Physica D} \textbf{\bibinfo{volume}{119}},
  \bibinfo{pages}{34} (\bibinfo{year}{1998}).

\bibitem[{\citenamefont{Cuevas et~al.}(2002{\natexlab{a}})\citenamefont{Cuevas,
  Archilla, Gaididei, and Romero}}]{CAGR02}
\bibinfo{author}{\bibfnamefont{J.}~\bibnamefont{Cuevas}},
  \bibinfo{author}{\bibfnamefont{J.}~\bibnamefont{Archilla}},
  \bibinfo{author}{\bibfnamefont{Y.}~\bibnamefont{Gaididei}}, \bibnamefont{and}
  \bibinfo{author}{\bibfnamefont{F.}~\bibnamefont{Romero}},
  \bibinfo{journal}{Physica D} \textbf{\bibinfo{volume}{163}},
  \bibinfo{pages}{106} (\bibinfo{year}{2002}{\natexlab{a}}).

\bibitem[{\citenamefont{Iba{\~n}es et~al.}(2002)\citenamefont{Iba{\~n}es,
  Sancho, and Tsironis}}]{IST02}
\bibinfo{author}{\bibfnamefont{M.}~\bibnamefont{Iba{\~n}es}},
  \bibinfo{author}{\bibfnamefont{J.}~\bibnamefont{Sancho}}, \bibnamefont{and}
  \bibinfo{author}{\bibfnamefont{G.}~\bibnamefont{Tsironis}},
  \bibinfo{journal}{Phys. Rev. E}  (\bibinfo{year}{2002}), \bibinfo{note}{in
  press}.

\bibitem[{\citenamefont{Cuevas et~al.}(2002{\natexlab{b}})\citenamefont{Cuevas,
  Palmero, Archilla, and Romero}}]{CPAR02}
\bibinfo{author}{\bibfnamefont{J.}~\bibnamefont{Cuevas}},
  \bibinfo{author}{\bibfnamefont{F.}~\bibnamefont{Palmero}},
  \bibinfo{author}{\bibfnamefont{J.}~\bibnamefont{Archilla}}, \bibnamefont{and}
  \bibinfo{author}{\bibfnamefont{F.}~\bibnamefont{Romero}},
  \bibinfo{journal}{Phys. Lett. A}  (\bibinfo{year}{2002}{\natexlab{b}}),
  \bibinfo{note}{submitted}.

\bibitem[{\citenamefont{Forinash et~al.}(1994)\citenamefont{Forinash, Peyrard,
  and Malomed}}]{FPM94}
\bibinfo{author}{\bibfnamefont{K.}~\bibnamefont{Forinash}},
  \bibinfo{author}{\bibfnamefont{M.}~\bibnamefont{Peyrard}}, \bibnamefont{and}
  \bibinfo{author}{\bibfnamefont{B.}~\bibnamefont{Malomed}},
  \bibinfo{journal}{Phys. Rev. E} \textbf{\bibinfo{volume}{49}},
  \bibinfo{pages}{3400} (\bibinfo{year}{1994}).

\bibitem[{\citenamefont{Archilla et~al.}(1999)\citenamefont{Archilla, MacKay,
  and Mar\'{\i}n}}]{AMM99}
\bibinfo{author}{\bibfnamefont{J.}~\bibnamefont{Archilla}},
  \bibinfo{author}{\bibfnamefont{R.}~\bibnamefont{MacKay}}, \bibnamefont{and}
  \bibinfo{author}{\bibfnamefont{J.}~\bibnamefont{Mar\'{\i}n}},
  \bibinfo{journal}{\mbox{Physica D}} \textbf{\bibinfo{volume}{134}},
  \bibinfo{pages}{406} (\bibinfo{year}{1999}).

\bibitem[{\citenamefont{Cuevas et~al.}(2001)\citenamefont{Cuevas, Archilla,
  Palmero, and Romero}}]{CAPR01}
\bibinfo{author}{\bibfnamefont{J.}~\bibnamefont{Cuevas}},
  \bibinfo{author}{\bibfnamefont{J.}~\bibnamefont{Archilla}},
  \bibinfo{author}{\bibfnamefont{F.}~\bibnamefont{Palmero}}, \bibnamefont{and}
  \bibinfo{author}{\bibfnamefont{F.}~\bibnamefont{Romero}},
  \bibinfo{journal}{J. Phys. A: Math. Gen.} \textbf{\bibinfo{volume}{34}},
  \bibinfo{pages}{L1} (\bibinfo{year}{2001}).

\bibitem[{\citenamefont{Peyrard and Bishop}(1989)}]{PB89}
\bibinfo{author}{\bibfnamefont{M.}~\bibnamefont{Peyrard}} \bibnamefont{and}
  \bibinfo{author}{\bibfnamefont{A.}~\bibnamefont{Bishop}},
  \bibinfo{journal}{Phys. Rev. Lett.} \textbf{\bibinfo{volume}{62}},
  \bibinfo{pages}{2755} (\bibinfo{year}{1989}).

\bibitem[{\citenamefont{Mar\'{\i}n and Aubry}(1996)}]{MA96}
\bibinfo{author}{\bibfnamefont{J.}~\bibnamefont{Mar\'{\i}n}} \bibnamefont{and}
  \bibinfo{author}{\bibfnamefont{S.}~\bibnamefont{Aubry}},
  \bibinfo{journal}{\mbox{Nonlinearity}} \textbf{\bibinfo{volume}{9}},
  \bibinfo{pages}{1501} (\bibinfo{year}{1996}).

\bibitem[{\citenamefont{Kopidakis and Aubry}(1999)}]{KA99}
\bibinfo{author}{\bibfnamefont{G.}~\bibnamefont{Kopidakis}} \bibnamefont{and}
  \bibinfo{author}{\bibfnamefont{S.}~\bibnamefont{Aubry}},
  \bibinfo{journal}{\mbox{Physica D}} \textbf{\bibinfo{volume}{130}},
  \bibinfo{pages}{155} (\bibinfo{year}{1999}).

\bibitem[{\citenamefont{Flach}(1996)}]{F96}
\bibinfo{author}{\bibfnamefont{S.}~\bibnamefont{Flach}},
  \bibinfo{journal}{Physica D} \textbf{\bibinfo{volume}{91}},
  \bibinfo{pages}{223} (\bibinfo{year}{1996}).

\bibitem[{\citenamefont{Ting and Peyrard}(1996)}]{TP96}
\bibinfo{author}{\bibfnamefont{J.}~\bibnamefont{Ting}} \bibnamefont{and}
  \bibinfo{author}{\bibfnamefont{M.}~\bibnamefont{Peyrard}},
  \bibinfo{journal}{Phys Rev E} \textbf{\bibinfo{volume}{53}},
  \bibinfo{pages}{1011} (\bibinfo{year}{1996}).

\bibitem[{\citenamefont{Bang and Peyrard}(1996)}]{BP96}
\bibinfo{author}{\bibfnamefont{O.}~\bibnamefont{Bang}} \bibnamefont{and}
  \bibinfo{author}{\bibfnamefont{M.}~\bibnamefont{Peyrard}},
  \bibinfo{journal}{Phys. Rev. E} \textbf{\bibinfo{volume}{53}},
  \bibinfo{pages}{4143} (\bibinfo{year}{1996}).

\bibitem[{\citenamefont{Salerno and Kivshar}(1994)}]{SK94}
\bibinfo{author}{\bibfnamefont{M.}~\bibnamefont{Salerno}} \bibnamefont{and}
  \bibinfo{author}{\bibfnamefont{Y.}~\bibnamefont{Kivshar}},
  \bibinfo{journal}{Phys. Lett. A} \textbf{\bibinfo{volume}{193}},
  \bibinfo{pages}{263} (\bibinfo{year}{1994}).

\end{thebibliography}
\end{document}